\documentclass[12pt]{article}

\usepackage{graphicx,isolatin1} 

\input babarsym
\renewcommand{\gg}{\ensuremath{\gamma\gamma}}

\begin{document}

~ \hfill \today

\begin{center} \huge \bf
Results on conventional and exotic charmonium at \babar
\end{center}

\begin{center} \large
Denis Bernard, ~
on behalf of the \babar\ Collaboration
\\
~
\\
Laboratoire Leprince-Ringuet, Ecole Polytechnique, CNRS/IN2P3, 
F-91128 Palaiseau, France 
\\
~
\\
Presented at 
XXI International Workshop on Deep-Inelastic Scattering and Related
Subject -DIS2013, 22-26 April 2013 Marseille, France 
\\ ~ \\ 

To appear in Proceedings of Science
\end{center}

{ \it
The \B\ factories provide a unique playground for studying the
properties of conventional and exotic charmonium states. We present
recent results in initial state radiation and two-photon fusion,
obtained using the full data set collected by the \babar\ experiment.
Amongst \babar\ 's harvest presented in this talk, the determination of
the quantum numbers of the X(3915) resonance, a body of concording
evidence pointing to $J^{PC} = 1^{++}$ for the X(3872), and updates on
the family of the Y resonance to the full integrated
luminosity. 
}

~

~

In this talk I have presented recent results obtained by the BaBar
experiment on charmonium and charmonium-like resonances, using
production mechanisms that select the quantum numbers.
Angular momentum conservation, and for strong decays parity
conservation and charge conjugation invariance then imply that these
quantum numbers also apply to the final state.
In the case of two-photon fusion, charge conjugation is positive, and
for non-tagged events, for which the initial electron and positron are
unobserved, the event selection performed during the analysis is such
that the two intermediate photons are quasi-real : 
allowed $J^{PC}$ are 
$0^{\pm+}, 2^{\pm+}, 4^{\pm+} \cdots$ and $3^{++}, 5^{++} \cdots$.
In the case of initial state radiation (ISR), for which a photon is
emitted by either of the incoming leptons, $J^{PC} = 1^{--}$. 

Results presented here are related to understanding the nature and the
properties of ``new'' resonances X(3872), X(3915), and the Y family.

\section{Study of X(3915) and search for X(3872), decaying to $\jpsi\omega$ in two-photon collisions}

Several charmonium-like states have been observed by the \B\ factories
in the mass region above the $D\overline{D}$ threshold, with
properties that disfavor their interpretation as conventional
charmonium mesons 
 \cite{Aubert:2005rm,Coan:2006rv,Yuan:2007sj,Aubert:2006ge,Wang:2007ea,BelleZ,BabarZ,Y3940a,Y3940b}.

The X(3915) resonance, decaying to the
$\jpsi\omega$ final state, was first observed by the Belle Collaboration
in two-photon collisions~\cite{BELLE_X3915}.
The $Z(3930)$ resonance was discovered in the $\gg\to
D\overline{D}$ process~\cite{BelleZ,BabarZ}. Its interpretation as the
$\chi_{c2}(2P)$, the first radial excitation of the $\phantom{}^3P_2$
charmonium ground state, is commonly accepted~\cite{PDG}. 
Interpretation of the X(3915) as the $\chi_{c0}(2P)$~\cite{Liu10} or
$\chi_{c2}(2P)$ state~\cite{Branz11} has been suggested. 
The latter would imply that the X(3915) and $Z(3930)$ are the same
particle, observed in different decay modes.

Despite the many measurements available~\cite{PDG}, the nature of the
$X(3872)$ state, which was first observed by Belle~\cite{BelleX},
is still unclear~\cite{Brambilla}.
The observation of its decay into $\gamma\jpsi$~\cite{X3872_rad}
ensures that this particle has positive $C$-parity.
The spin analysis performed by CDF on the decay
$X(3872)\to\jpsi\pip\pim$ concludes that only $J^P=1^{+}$ and
$J^P=2^{-}$ are consistent with data~\cite{X3872_cdf}.
If $J^P=2^-$, the production of the $X(3872)$ in two-photon collisions
would be allowed, while for $J=1$ if would be forbiden.

\babar\ searched for the X(3872) and the X(3915) resonances in the
decay mode $\jpsi\omega$ and studied the quantum numbers of the
X(3915) with angular analyses \cite{Lees:2012xs}, in a sample of
events corresponding to 519.2 \invfb.
Intermediate mesons are reconstructed in the modes $\jpsi \to
\ell^+\ell^-, \ (\ell = e\mbox{ or }\mu$) and $\omega \to \pi^+ \pi^-
\pi^0$.
Besides the usual event reconstruction and signal selection, 
several cuts are applied to reject specific backgrounds.
Events with extra tracks with momentum greater than 0.1 \gevc,
or with a large energy non-associated with the event ($E_{extra}>0.3\gev$), 
 events from ISR $\jpsi \pi^+ \pi^- \pi^0$ production, and 
residual ISR $\psitwos \to \jpsi \pi^+ \pi^- $ production are vetoed.
The \piz\ mass is constrained to its nominal value~\cite{PDG}.
To improve the mass resolution, we define the $\jpsi\omega$ mass as
$m(\jpsi\omega) = 
m(\ell^+\ell^-\pi^+ \pi^- \pi^0) - m(\ell^+\ell^-) + m(\jpsi)^{\mathrm{PDG}}$. 

\begin{figure}[!h]
 \begin{center}
  \includegraphics[scale=0.5]{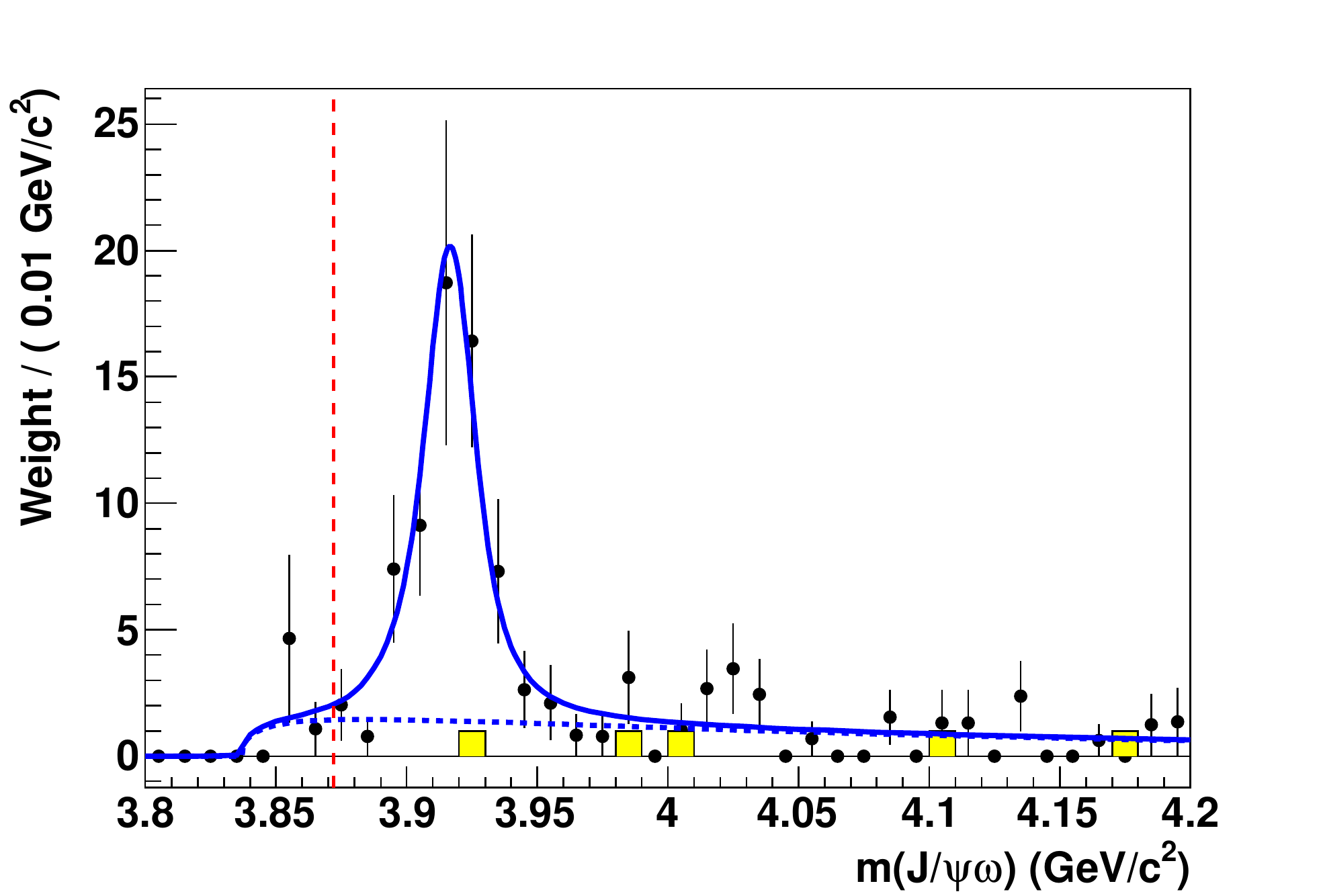}
  \caption{The efficiency-corrected $m(\jpsi\omega)$ distribution
  (solid points) \cite{Lees:2012xs}. 
The shaded histogram is the non-$\jpsi\omega$ background
  estimated from sidebands. The vertical dashed (red) line is
  at $m(\jpsi\omega)=3.872~\gevcc$.}
 \label{fig:fit}
 \end{center}
\end{figure}

\begin{figure}[!h]
 \begin{center}
  \includegraphics[scale=0.34]{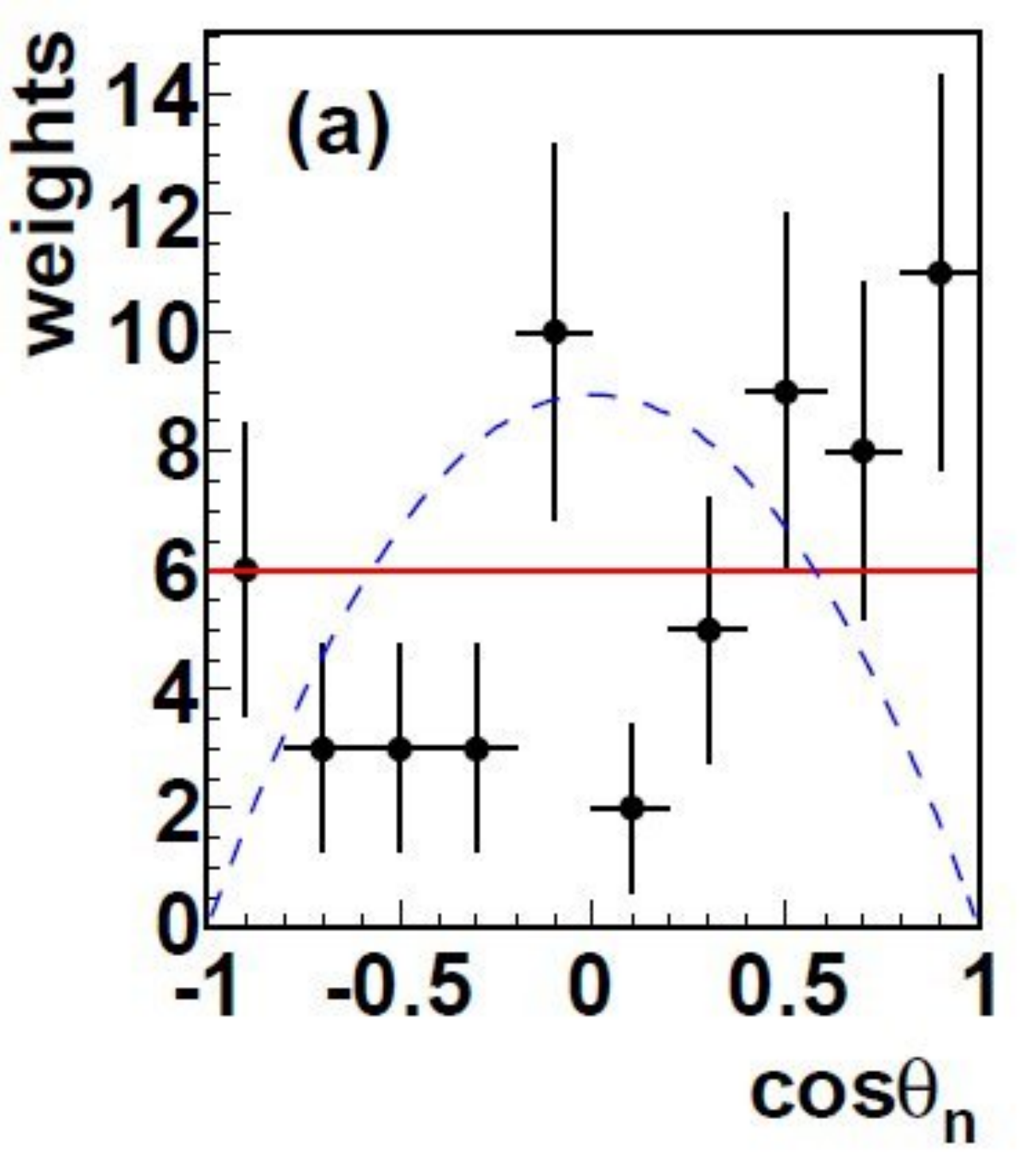}
  \includegraphics[scale=0.34]{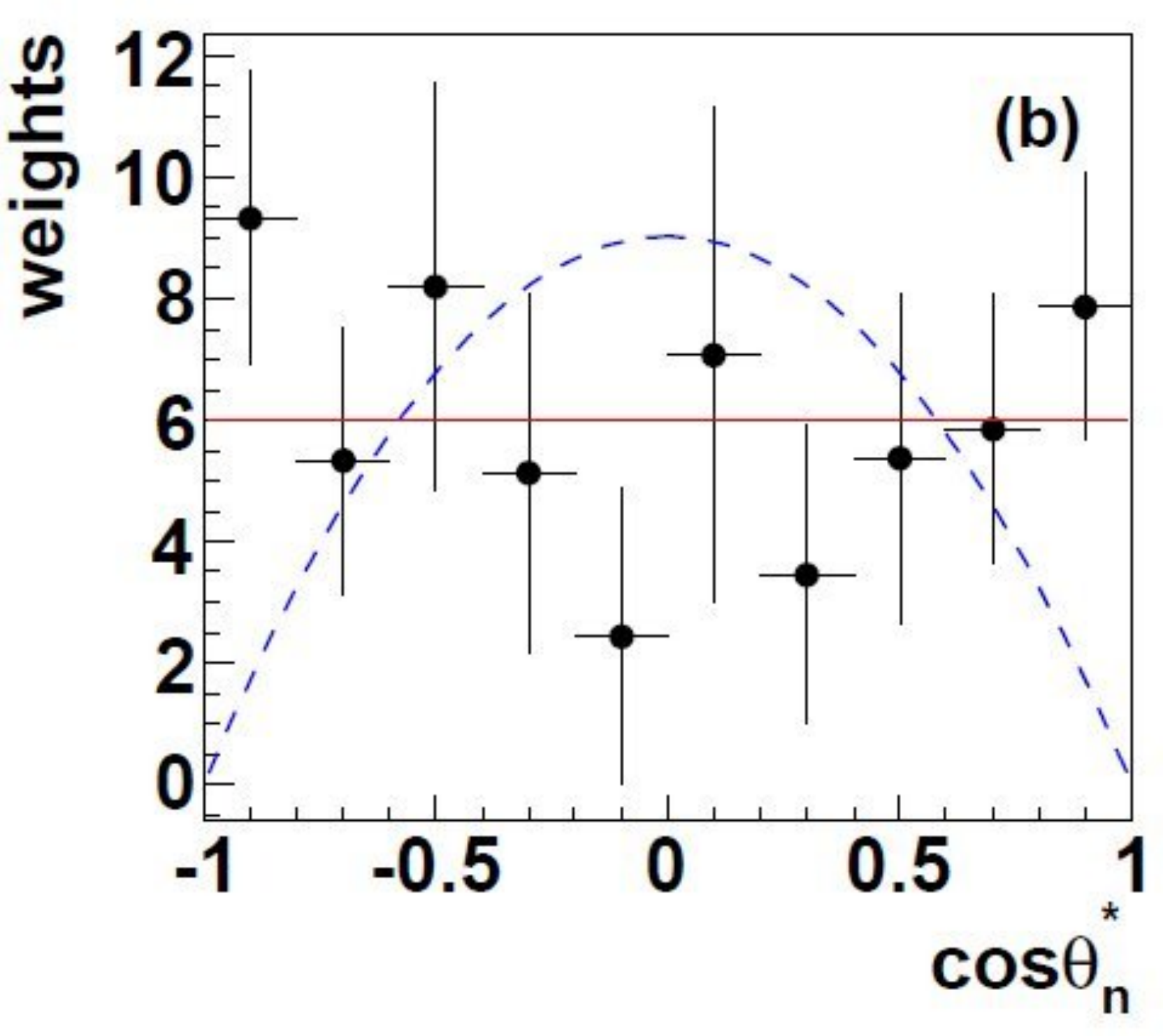}
  \caption{The efficiency-corrected distributions of
  $\cos{\theta_n}$ and $\cos{\theta_n^*}$ for events in the
  X(3915) mass signal region \cite{Lees:2012xs}. 
 \label{fig:ang}}
 \end{center}
\end{figure}

We observe a prominent peak near $3915~\mevcc$ over a small background
(Fig.~\ref{fig:fit}), with $59\pm10$ signal events, a mass of
$(3919.4\pm 2.2\pm 1.6)~\mevcc$ and a width of $(13\pm 6 \pm 3)~\mev$
with a significance of $7.6 \sigma$, results which are compatible
with those of Belle ~\cite{BELLE_X3915}.
No structure is seen around $3872~\mevcc$, with an upper limit of 
$\Gamma_{\gg}[X(3872)]\times\calB(X(3872)\to\jpsi\omega) < 1.7~\ev$ 
at 90\% CL, assuming $J=2$.
We first discriminate between $J^P=0^{\pm}$ and $J^P=2^+$ by using the
Rosner~\cite{Rosner} predictions.
The distribution of several angles that determine the final state are
examined.
In all cases the $J^P=0^{\pm}$ hypothesis describes the data better
than the $J^P=2^+$ \cite{Lees:2012xs}.
For the distribution of the angle $\theta_n^*$ between the normal
to the decay plane of the $\omega$ and the two-photon axis
(Fig. \ref{fig:ang} right)
 $\chi^2$ probabilities for $J^P=0^{\pm}$ and $J^P=2^+$
are respectively 64.7\% and $9.6\times 10^{-9}$\% respectively. 
We then discriminate between $J^P=0^-$ and $J^P=0^+$. In all cases
the $J^P=0^+$ hypothesis gives a smaller $\chi^2$ than the $J^P=0^-$
hypothesis.
For the distribution of the angle $\theta_n$ between the normal to the
$\omega$ decay plane and the $\omega$ direction in the $\jpsi \omega$
rest frame
(Fig. \ref{fig:ang} left)
 $\chi^2$ probabilities for $J^P=0^+$ and $J^P=0^-$ are
6.1\% and 4.8$\times10^{-11}$\% respectively.
We find that $J^P=0^{\pm}$ is largely preferred over $J^P=2^+$ and
that $J^P=0^+$ is largely preferred $J^P=0^-$, which points to an
interpretation of the X(3915) resonance being the $\chi_{c0}(2P)$
charmonium meson.

\section{Search for resonances decaying to $\eta_c \pi^+ \pi^-$ in two-photon interactions}

An other final state that may be considered to test X(3872) production
in two-photon, still with the hope to discriminate $J^{PC} = 1^{++}$
from $J^{PC} = 2^{-+}$, is $\eta_c \pi^+ \pi^-$.
Voloshin ~\cite{Voloshin:2002xh} compared the $\eta_c' \to \eta_c$
and $\psitwos \to \jpsi$ dipion transitions in terms of QCD multipole
expansion, and predicted 
$\BR(\eta_c(2S) \to \eta_c\pip\pim) / \BR(\psi(2S) \to
\jpsi\pip\pim) = 2.9$, that is 
$\BR(\eta_c(2S) \to \eta_c\pip\pim) = (2.2^{+1.6}_{-0.6})\%$.

In the same spirit, with the possible assignment of the X(3872) to the
charmonium meson $\eta_{c2}$ ($1^1D_2$, with
$J^{PC}=2^{-+}$), the branching fraction of the isospin-conserving
decay $X(3872) \to \eta_c \pip\pim$ would be larger than that of the
isospin-violating decay discovery channel $X(3872) \to \jpsi\pip\pim$
\cite{Olsen:2004fp}, which is larger than $2.6\,\%$\cite{PDG}.
Two photon production would be allowed, and the event rate could be
sizable.

The \babar\ experiment studied resonances decaying to $\eta_c \pi^+
\pi^-$ in two-photon interactions, using a data sample that
corresponds to 473.9 \invfb \cite{Lees:2012me}.
$\eta_c$ candidates were reconstructed in the $\KS \Kp \pim$ decay
mode.
Background rejection is performed in a very sophisticated way.
After an usual event reconstruction and a two-photon-like preselection
of the candidates, cuts are applied in the $\eta_c$ decay Dalitz plane
to take advantage of the fact that $\eta_c\to \KS \Kp \pim$ decays
often proceed via intermediate $K^{*}_{0}(1430)$ states, while
background events often contain a $K^*(892)$ meson.
Finally a cut is applied on the output of a neural network whose
inputs include general properties of the $\eta_c\pip\pim$ candidate
such as $p_T$ and $E_{extra}$, and particle identification 
information for the charged tracks.

 \begin{figure}[!h]
 \begin{center}
  \includegraphics[scale=0.24]{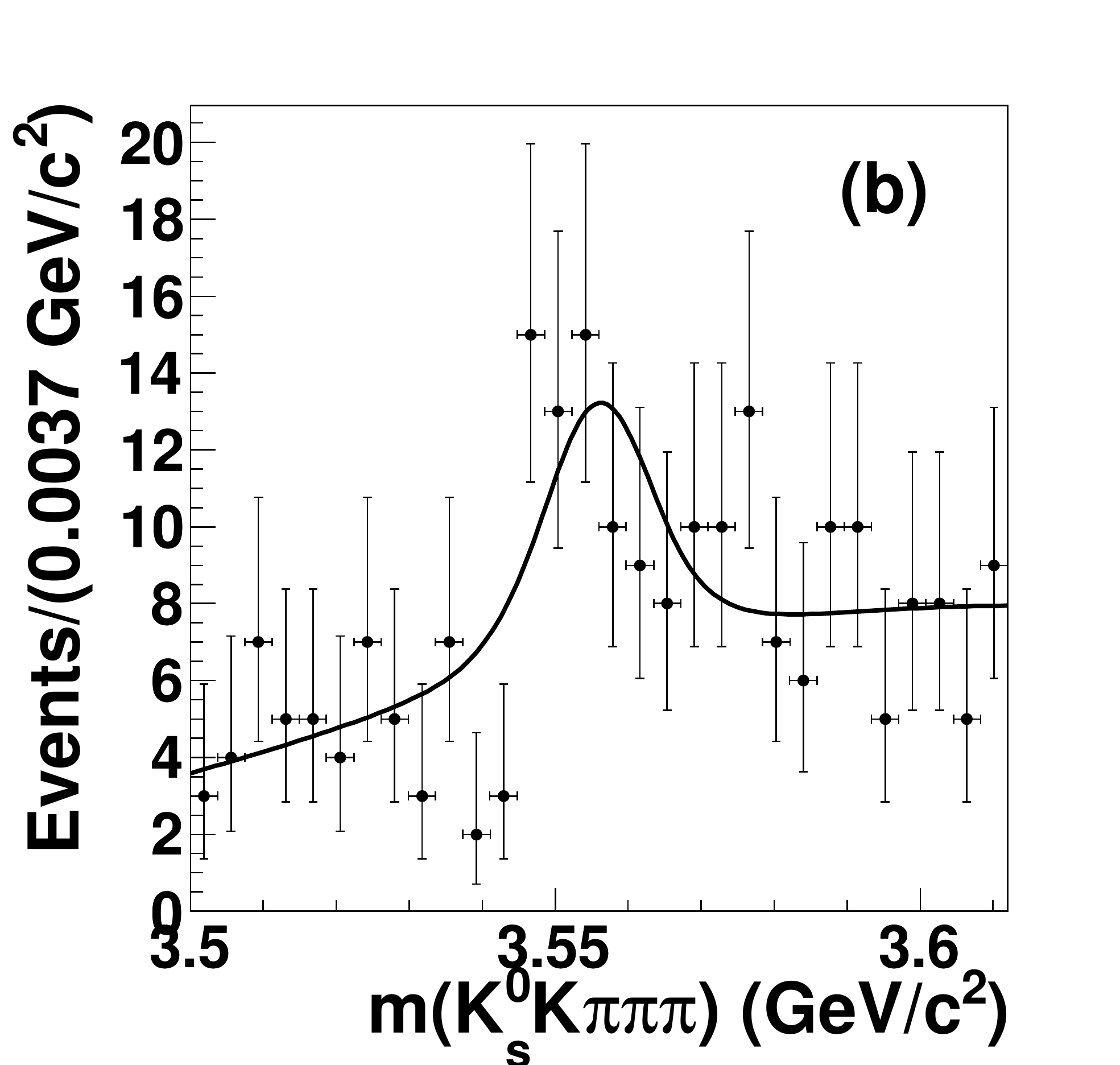}
  \includegraphics[scale=0.24]{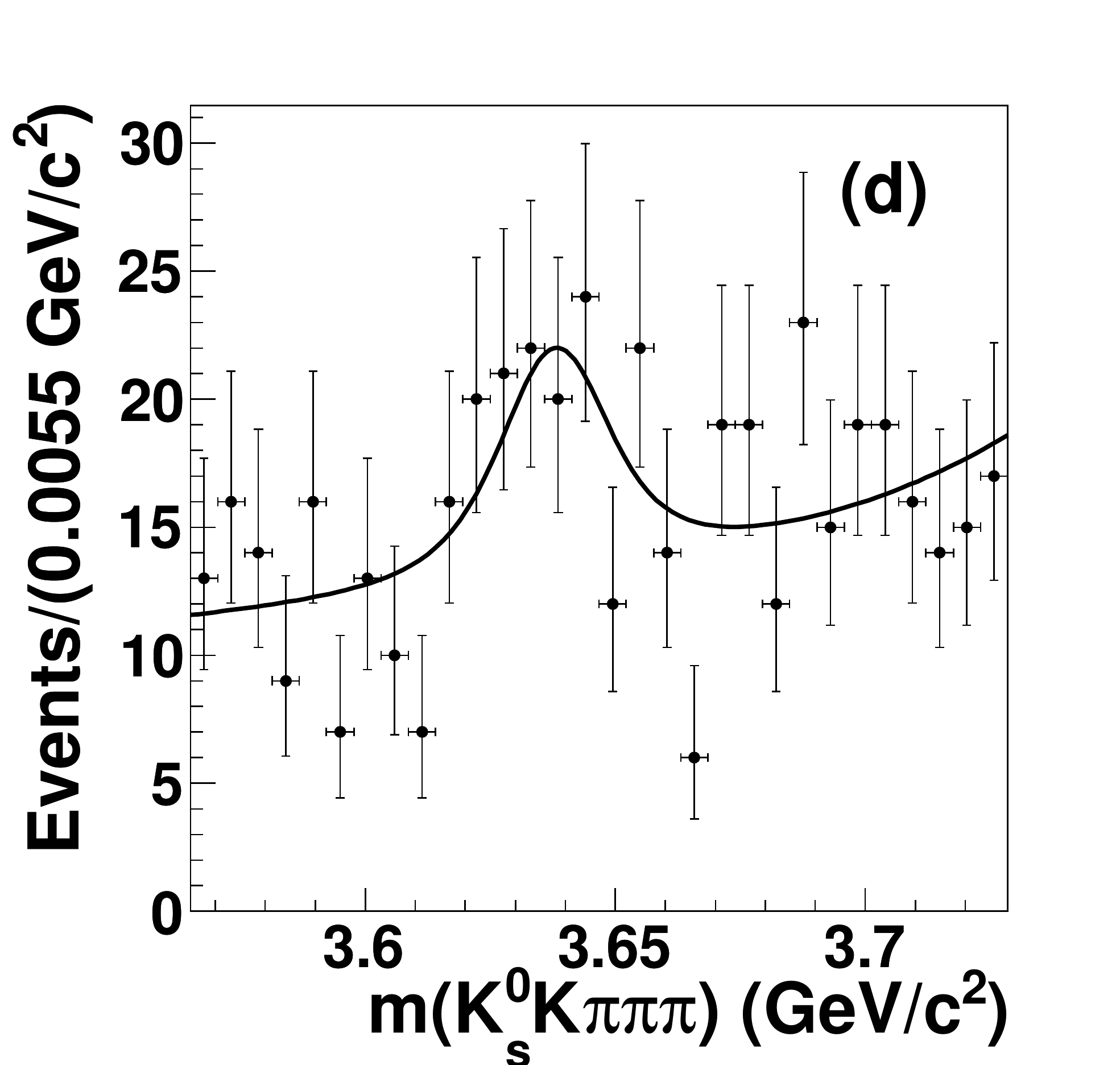}
  \includegraphics[scale=0.24]{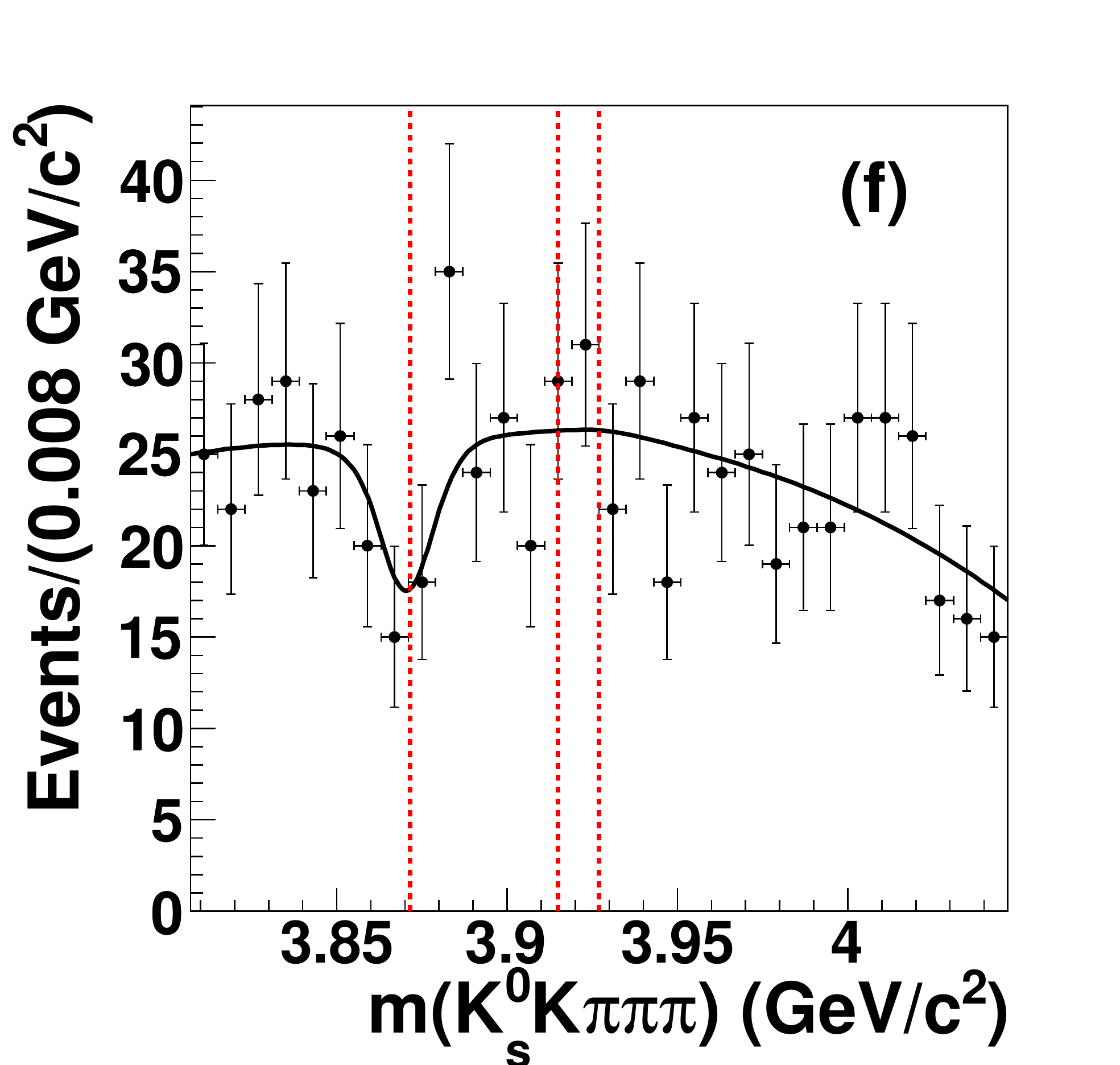}
  \caption{Distributions of 
$m_{\KS \Kp \pim \pip \pim}$ in three mass ranges of interest, corresponding to the 
 $\chi_{c2}(1P)$, to the $\etac(2S)$, and to the $X(3872)$, $X(3915)$ and $\chi_{c2}(2P)$ (the masses of which are indicated by the vertical lines)
 \cite{Lees:2012me}.
 \label{fig:etacpipi}}
 \end{center}
 \end{figure}

Two strutures are found, for masses that are compatible with that of
the $\chi_{c2}(1P)$ and of the $\etac(2S)$ meson, but no sign of any
signal is detected for the $X(3872)$, $X(3915)$ and $\chi_{c2}(2P)$
(Fig. \ref{fig:etacpipi})
which indicates again that the $2^{-+}$ assignment is disfavored for
the $X(3872)$.
We obtain an upper limit $\BR(\eta_c(2S) \to \eta_c \pip\pim) < 7.4\,\%$ 
that is compatible with the prediction of 2.2\%.

\section{Study of ISR-produced resonances decaying to $\psi \pip \pim$}

The Y(4260) resonance discovered by \babar\ \cite{Aubert:2005rm} in ISR
production to $\jpsi \pip \pim$ final state, has escaped firm
intepretation to date.
Most likely is it not a regular charmonium meson since all
attempts to detect its decays to either $D \Db$, $D \Dstarb$,
$\Dstar\Dstarb$, $D_s^+D_s^-$, $D_s^{*+}D_s^{-}$, $D_s^{*+}D_s^{*-}$
have failed.
Most likely it is not a four-quark system since the most natural
assignment would have it decay dominantly to $D_s^+D_s^-$
\cite{Maiani:2005pe}, which are not seen.
Searching for Y(4260) decays to $\psitwos \pip \pim$, \babar\ didn't
find any, but found instead a new similar resonance at an invariant
mass of 4.32 \gevcc \cite{Aubert:2006ge}, now named the the Y(4330).
Belle confirmed the existence of the Y(4260) resonance,
claimed a hint of an other object at a lower mass, close to 4 \gevcc, 
confirmed the existence of the Y(4330) resonance, 
and discovered a new resonance, the Y(4660), in the same channel
$\psitwos \pip \pim$ \cite{Wang:2007ea}.

We report here a recent update by \babar\ of the 
 $\jpsi \pip \pim$ analysis to 454 \invfb 
\cite{Lees:2012cn}, and the first \babar\ study of the 
 $\psitwos \pip \pim$ final state, using 520 \invfb 
\cite{Lees:2012pv}.
\begin{figure}[!h]
 \begin{center}
 \setlength{\unitlength}{0.47\textwidth}
 \begin{picture}(1,0.75)(0,0)
 \put(0,0){
\includegraphics[width=0.47\textwidth]{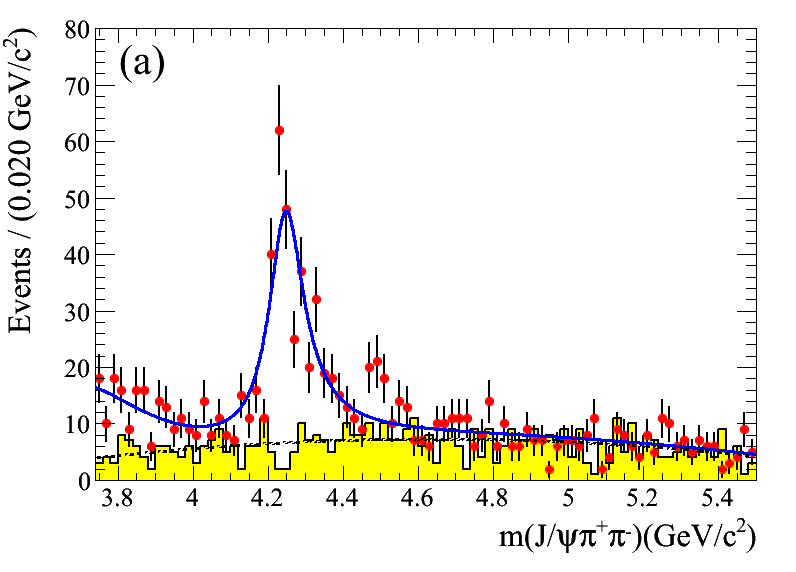}
}
 \end{picture}
\setlength{\unitlength}{0.47\textwidth}
 \begin{picture}(1,0.75)(0,0)
 \put(0,0){
\includegraphics[width=0.47\textwidth]{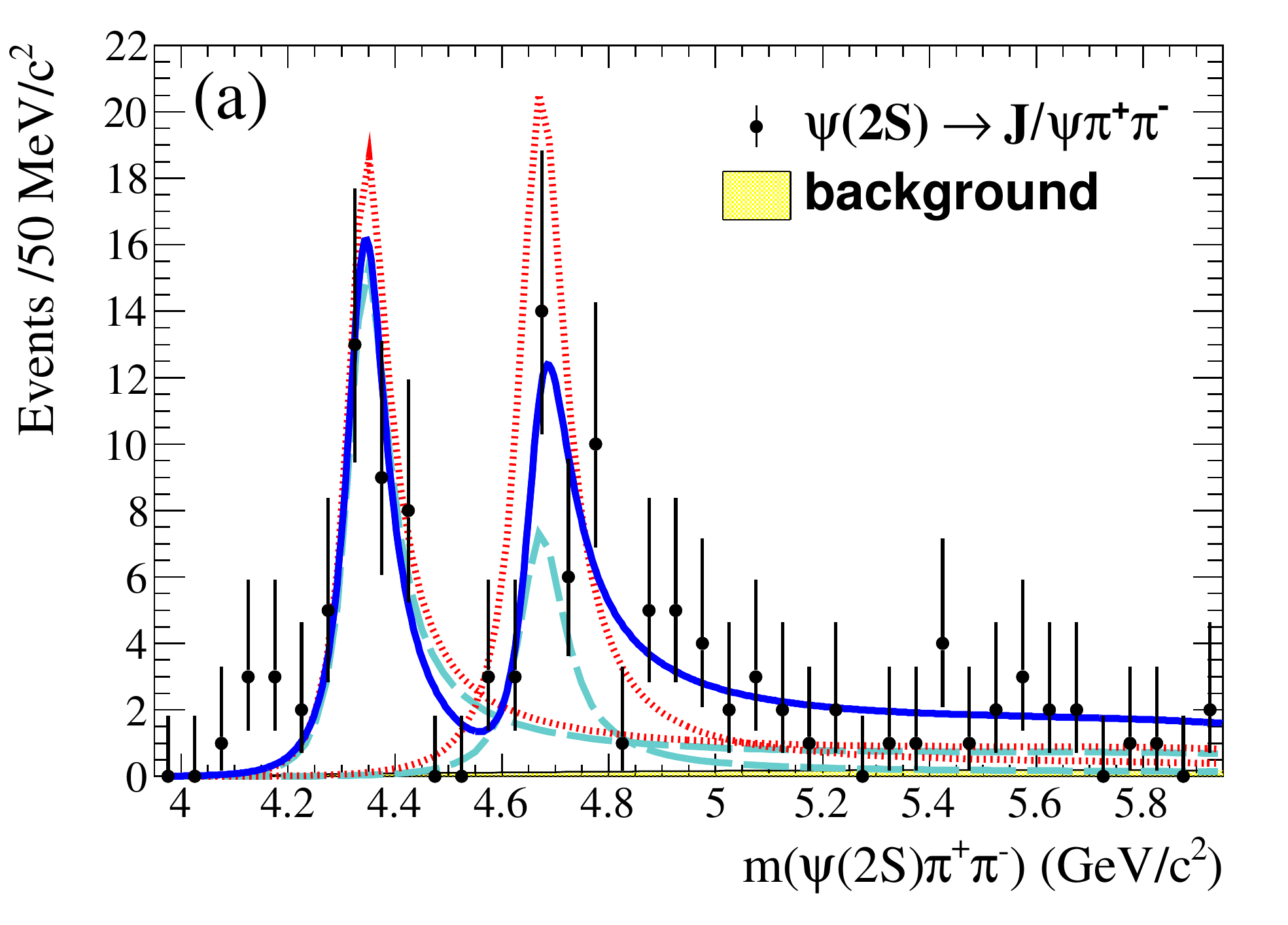}}
 \put(0.55,0.42){{\scriptsize \babar\ Preliminary}}
 \end{picture}
  \caption{Mass distributions of the 
 $\jpsi \pip \pim$ (left, \cite{Lees:2012cn}) and
 $\psitwos \pip \pim$ (right, \cite{Lees:2012pv})
final states.
 \label{fig:psipipi}}
 \end{center}
\end{figure}
\babar\ confirms (\cite{Lees:2012cn}, Fig. \ref{fig:psipipi} left) 
its discovery of the ISR-produced
$Y(4260) \to \jpsi \pip \pim$ \cite{Aubert:2005rm} and Belle's
observation of $Y(4360)$ and $Y(4660)$ to $\psitwos \pip \pim$, but
does not see any sign of an $Y(4008) \to \jpsi \pip \pim$
 (\cite{Lees:2012pv}, Fig. \ref{fig:psipipi} right). 
I agree with the Belle speaker \cite{Santel} that this thing, which
was presented without a significance, should be handled with quotes.

\section{Conclusion}

We have presented studies of untagged production in two-photon fusion
with the $\jpsi\omega$ and $\eta_c \pip \pim$ final states.
Together with its two-photon production, angular analyses of $X(3915)
\to \jpsi\omega$ determine the quantum numbers of the $X(3915)$ to be
$J^{PC} = 0^{++}$, indicating the possible assignment to a
$\chi_{c0}(2P)$ charmonium meson.
The searches of $X(3872)$ in these two final states don't show any
significant signal, despite at least the branching fraction of
$X(3872)$ to $\eta_c \pip \pim$ is expected to be sizable in the case
of the $J^{PC} = 2^{-+}$.
These two results, together with LHCb's preliminary angular analysis
of $X(3872)\to \jpsi \pip \pim$ \cite{Aaij:2013zoa} point to a 
$J^{PC} = 1^{++}$ assignment for the $X(3872)$.
The proximity of its mass to the $\Dstarz\Dzb$ threshold suggested it to be a 
$J^{PC} = 1^{++}$ hadronic molecule \cite{Tornqvist:2004qy} or a 
a tetraquark \cite{Maiani:2004vq}.
The family of the $J^{PC} = 1^{--}$ Y states produced in ISR with
$\psi\pip\pim$ is now well established, with 3 confirmed states. We
may be seeing the dawn of a hybrid charmonium spectroscopy
\cite{hybrides}.

\end{document}